\documentclass[aps,showpacs,preprintnumbers,amsmath, amssymb]{revtex4}

\usepackage{float}
\usepackage{graphics,epsfig}
\usepackage{graphicx}
\usepackage{dcolumn}
\usepackage{bm}

\begin{document}
\baselineskip=0.8 cm
\title{{\bf The torsion cosmology in Kaluza-Klein theory}}
\author{Songbai Chen}
\email{csb3752@163.com} \affiliation{Institute of Physics and
Department of Physics,
Hunan Normal University,  Changsha, Hunan 410081, P. R. China \\
Key Laboratory of Low Dimensional Quantum Structures and Quantum
Control (Hunan Normal University), Ministry of Education, P. R.
China.}

\author{Jiliang Jing }
\email{jljing@hunnu.edu.cn}
 \affiliation{Institute of Physics and
Department of Physics,
Hunan Normal University,  Changsha, Hunan 410081, P. R. China \\
Key Laboratory of Low Dimensional Quantum Structures and Quantum
Control (Hunan Normal University), Ministry of Education, P. R.
China.}

\vspace*{0.2cm}
\begin{abstract}
\baselineskip=0.6 cm
\begin{center}
{\bf Abstract}
\end{center}

We have studied the torsion cosmology model in Kaluza-Klein theory.
We considered two simple models in which the torsion vectors are
$A_{\mu}=(\alpha,0,0,0)$ and $A_{\mu}=a(t)^2(0,\beta,\beta,\beta)$,
respectively. For the first model, the accelerating expansion of the
Universe can be not explained without dark energy which is similar
to that in the standard cosmology. But for the second model, we find
that without dark energy the effect of torsion can give rise to the
accelerating expansion of the universe and the alleviation of the
well-known age problem of the three old objects for appropriated
value of the model parameter $\beta$. These outstanding features of
the second torsion cosmology model have been supported by the Type
Ia supernovae (SNIa) data.

\end{abstract}

\maketitle
\newpage
\vspace*{0.2cm}

\section{Introduction}

Many observations \cite{A1,A2,A3} has been strongly confirmed that
the expansion of our present Universe is accelerating rather than
slowing down. This late time cosmic acceleration can not be
explained by the four known fundamental interactions in the standard
models, which is the greatest challenge today in the modern physics.
Within the framework of Einstein's general relativity, an exotic
component with negative pressure called dark energy is invoked to
explain this observed phenomena. The simple candidate of dark energy
which is consistent with current observations is the cosmological
constant \cite{1a}, which is a term that can be added to Einstein's
equations. This term acts like a perfect fluid with an equation of
state $\omega=-1$, and the energy density is associated with quantum
vacuum. However, it is very difficult to understand in the modern
field theory since the vacuum energy density is far below the value
predicted by any sensible quantum field theory. Moreover, it has
also been plagued by the so-called coincidence problem. Thus, a lot
of the dynamical scalar fields, such as quintessence \cite{2a},
k-essence \cite{3a}, phantom \cite{4a} and quintom field
\cite{5a,5a1}, have been put forth as an alternative of dark energy.
However, so far, the nature of dark energy is still unclear.

On the other hand, it is argued that in Einstein theories gravity is
not well understood and an important ingredient is missing which may
account for such observed phenomena. One of such an ingredient is
the torsion which is vanished in Einstein theories. It is widely
believed that the presence of torsion will change the character of
the gravitational interaction because that in this case the
gravitational field is described not only by spacetime metric, but
also by the torsion field. Recently, a lot of investigations have
indicated that the torsion plays the important role in the modern
physics \cite{tor1}. Cosmological models with torsion were pioneered
by Kopczy\'{n}ski \cite{tor2} in the last century. Thereafter, the
bouncing cosmological model with torsion has been proposed by
Kerlick \cite{tor3} in which the torsion was imagined as playing
role only at high densities in the early universe. The effects of
torsion field on the inflation in cosmology has been investigated in
\cite{tor4,tor5}. Recently some authors have begun to study torsion
as a possible reason of the accelerating universe \cite{tor6}. The
study of dynamics and the statefinder diagnostic in torsion
cosmology have been studied in \cite{Li1,Li2}.

Recently, the cosmology in Kaluza-Klein theory
\cite{KKlein1,KKlein2,KKlein3} has attracted a considerable
attention \cite{KKC1,KKC2,KKC3,KKC4,Cho1,Cho2,Cokk,Cokk1,KKDE}. The
Kaluza-Klein theory was introduced first by Kaluza \cite{KKlein1} to
unify Maxwell's theory of electromagnetic and Einstein's gravity
theory by adding the fifth dimension. In 1926, Klein
\cite{KKlein2,KKlein3} proposed that the fifth dimension is
compactified by being curled up in a circle of very small radius, so
that the extra dimension is not observable except on very high
energy scales. Due to its potential function to unify the
fundamental interactions, Kaluza-Klein theory has been regarded as a
candidate of fundamental theory. Moreover, the fantastic idea of
extra dimensions promotes various higher dimensional theories,
including the well-known string theory \cite{String}. Thus
Kaluza-Klein theory has been revived in recent years in the modern
physics, such as in supergravity \cite{supergr} and superstring
theories \cite{superst}. Many of the papers currently published in
the context of Kaluza-Klein theory deal with cosmology. Cho
\cite{Cho1,Cho2} has studied the $4+1$ dimensional Kaluza-Klein
cosmology and find that it can solve some of the problems in the
standard big bang model (including the horizon problem and the
flatness problem ) in a very natural way. Wesson \textit{et al}
\cite{Cokk} have investigated the cosmological constant problem in
Kaluza-Klein cosmology. Li \cite{Cokk1} has considered the inflation
in Kaluza-Klein theory and obtain a relation between the
fine-structure constant and the cosmological constant. Some authors
argued that Kaluza-Klein theory can be effective in accounting for
the dark constituent of the universe \cite{KKDE}. In Ref.
\cite{KK1}, Shankar \textit{et al} incorporates the torsion into the
five dimensional Kaluza-Klein theory \cite{KKlein1,KKlein2,KKlein3}
and find that there exists some non-vacuum solutions. Moreover, they
discuss furtherly the implications of Kaluza-Klein theory with
torsion fields in cosmology and obtain that the presence of torsion
fields changes the dynamical equations for a flat
Friedmann-Robertson-Walker spacetime. This means that the torsion
fields will play the important roles in the evolution of the
Universe. The motivation of this paper is to study concretely the
properties of the torsion cosmology in the Kakluza-Klein theory and
then perform the constraints on this cosmology model by using the
Type Ia supernovae (SNIa) data.

The remainder of this paper is organized as follows: in the next
section we review briefly the torsion cosmology in Kaluza-Klein
theory. And then we study the cosmic expansion history and check the
age problem of the high redshift objects for the models. In Sec.III,
we perform the constraints on this cosmology model by using the SNIa
data and present our results. Finally in the last section we will
include our conclusions.

\section{The torsion cosmology in Kaluza-Klein theory }

In this section, we first review briefly the torsion cosmology in
Kaluza-Klein theory which proposed by Shankar \textit{et al}
\cite{KK1}, where the torsion tensor $S^c_{\;ab}$ is defined as the
antisymmetric part of connection in a coordinate basis
\begin{eqnarray}
S^{i}_{\;jk}=\Gamma^{i}_{\;jk}-\Gamma^{i}_{\;kj},
\end{eqnarray}
where $\Gamma^{i}_{\;jk}$ is the affine connection with torsion and
has a form
\begin{eqnarray}
\Gamma^{i}_{\;jk}=\hat{\Gamma}^{i}_{\;jk}+K^{i}_{\;jk}.
\end{eqnarray}
The $\hat{\Gamma}^{i}_{\;jk}$ is the usual Christoffel symbol and
the $K^{i}_{\;kj}$ is the well-known contorsion tensor
\begin{eqnarray}
K^{i}_{\;jk}=\frac{1}{2}[S^{i}_{\;kj}+S^{\;i}_{k\;j}+S^{\;i}_{j\;k}].
\end{eqnarray}
Using some constraints and conditions, Shankar \textit{et al}
\cite{KK1} found the five dimensional Einstein equations in the
Kaluza-Klein theory
\begin{eqnarray}
G_{ij}=\tilde{R}_{ij}+\frac{1}{2}\mathbf{g}_{ij}\tilde{R}=\tilde{T}_{ij},
\end{eqnarray}
can be split into
\begin{eqnarray}
&&R_{\mu\nu}-\frac{1}{2}g_{\mu\nu}R-\frac{1}{2}A_{\mu}A_{\nu}\Phi^2R=T_{\mu\nu},\nonumber\\
&&-\frac{1}{2}A_{\mu}\Phi^2R=T_{\mu5},\;\;\;\;\;\;\;-\frac{1}{2}\Phi^2R=T_{55}.\label{g1}
\end{eqnarray}
Here $\mathbf{g}_{ij}$, $\tilde{R}_{ij}$ and $\tilde{T}_{ij}$ are
the five dimensional metric, Ricci curvature tensor and the
energy-momentum tensor in Kaluza-Klein theory respectively. The
$g_{\mu\nu}$, $R_{\mu\nu}$ and $T_{\mu\nu}$ are corresponding
tensors in the four dimensional hypersurface. $A_{\mu}$ and $\Phi$
are vector and scalar torsion fields. For the vacuum solution with
$\tilde{T}_{ij} = 0$ in 5D spacetime, one can obtain from Eq.
(\ref{g1}) that the Ricci scalar $R =0$ and $R_{\mu\nu}=0$, which
means that the 4D metric $g_{\mu\nu}$ obtained from vacuum solutions
in 5D spacetime are identical to the vacuum solutions in the torsion
free 4D spacetime \cite{KK1}. The converse is also true. Therefore,
for the vacuum solutions, we can not distinguish the 4D spacetime
from the 4D hypersurface within the 5D spacetime with torsion
\cite{KK1}. In other word, the effects of torsion can not be
detected in this case. However, for the non-vacuum solution, one can
construct new 4D metric $g_{\mu\nu}$ which do not exist in torsion
free 4D spacetime \cite{KK1}. Comparing with the solutions in
Einstein's theory, we can acquire some information about torsion.

In order to develop our cosmological considerations, let us take
into account a flat Friedmann-Robertson-Walker metric of the type
\begin{eqnarray}
ds^2=-dt^2+a(t)^2(dr^2+r^2d\theta^2+r^2\sin(\theta)^2d\phi^2).\label{me1}
\end{eqnarray}
Inserting the metric (\ref{me1}) into Eq.(\ref{g1}), we find that
due to the presence of the torsion field $A_{\mu}$ and $\Phi$ the
Friedmann equation needs to make some modifications. Thus comparing
the case without torsion, we find that the evolution of the Universe
will present the different properties. In general, the torsion
scalar $\Phi$ and vector fields $A_{\mu}$ vary with the time $t$.
The dynamical torsion scalar fields in cosmology have been studied
extensively in other theories \cite{tor6,Li1,Li2}. For simplicity,
in this paper we adopt $\Phi^2=1$ and consider the two simplest
models \cite{KK1} $A_{\mu}=(\alpha,0,0,0)$ and
$A_{\mu}=a(t)(0,\beta,\beta,\beta)$ and then study their cosmic
expansion histories, where $\alpha$ and $\beta$ are constants. These
torsion fields ansatze yield that the modified Friedmann equation
possesses a more simple form, which is very convenient for us to
study the dynamical properties of the model in the following
calculations. In order to probe of the properties of the torsion
cosmology in Kaluza-Klein theory, here we assume that the Universe
contains only the dark matter.

Let us now consider the first model $A_{\mu}=(\alpha,0,0,0)$ and
$\Phi^2=1$. Combining Eqs.(\ref{me1}) and (\ref{g1}), we find that
the modified Friedmann equation and Raychaudhuri field equationcan
be expressed simply as \cite{KK1}
\begin{eqnarray}
H^2=\frac{2}{3(2-\alpha^2)}\rho,\label{F1}
\end{eqnarray}
 and
\begin{eqnarray}
\dot{H}=-\frac{1}{2-\alpha^2}\rho,\label{FH1}
\end{eqnarray}
respectively, where $\rho$ is the energy density of the fluid. The
conservation law of the total energy reads
\begin{eqnarray}
\dot{\rho}+3H\rho=0,\label{E1}
\end{eqnarray}
For the dark matter $\omega=0$, the Eq. (\ref{E1}) are independent
of the constant $\alpha$, which are consistent with those in the
standard Einstein cosmology. The factor $2/(2-\alpha^2)$ in Eqs.
(\ref{F1}) and (\ref{FH1}) is equivalent to the constant of gravity.
The deceleration parameter
\begin{eqnarray}
q\equiv-\frac{\ddot{a}a}{\dot{a}^2}=-\frac{\dot{H}+H^2}{H^2}=\frac{1}{2}.
\end{eqnarray}
This means that in the first model the expansion is not accelerating
if the Universe contains only the dark matter. Thus, in order to
explain the accelerating expansion, we must introduce to the exotic
components as in the Einstein cosmology, such as dark energy.
Therefore, in the following section we do not consider further the
first model.

For the second model $A_{\mu}=a(t)(0,\beta,\beta,\beta)$ and
$\Phi^2=1$ \cite{KK1}, repeating the operations above, we obtain
that the Friedmann equation, Raychaudhuri field equation and the
conservation law of the total energy can be expressed as \cite{KK1}
\begin{eqnarray}
&&H^2=\frac{1}{3}\rho,\;\;\;\;\;\;\;\;\;\;\; \dot{H}=-\frac{1-2\beta^2}{2-3\beta^2}\rho,\label{FH2}\\
&&\dot{\rho}+3H\bigg[\frac{2-4\beta^2}{2-3\beta^2}\bigg]\rho=0.\label{E2}
\end{eqnarray}
In this case, the Friedmann equation is the same as in the torsion
free Einstein cosmology. But both the Raychaudhuri field and the
continue equations depend on the torsion parameter $\beta$, which is
different from that in the model we discuss previously. The
deceleration parameter is
\begin{eqnarray}
q\equiv-\frac{\ddot{a}a}{\dot{a}^2}=-\frac{\dot{H}+H^2}{H^2}=\frac{1-3\beta^2}{2-3\beta^2}.
\end{eqnarray}
Obviously, $q$ is a constant which is determined only by the torsion
parameter $\beta$. When $\beta^2$ lies in the region
$(\frac{1}{3},\frac{2}{3})$, we find that $q<0$. This means that the
expansion of the Universe can be accelerated without dark energy,
which is different from that in the Einstein cosmology.

Let us now to examine the second model with some old high redshift
objects including two old galaxies LBDS $53W091$ $(z=1.55, t=3.5
Gyr)$ \cite{ag1}, LBDS 53W069 ($z=1.43,t=4.0 Gyr)$ \cite{ag2} and
the old quasar APM 08279+5255 $(z=3.91, t=2.1 Gyr)$ \cite{ag3}. The
absolute ages of these oldest galactic globular clusters and quasars
provide a fundamental constraint on cosmological models because that
the universe cannot be younger than its constituents at any
redshift. The age of the Universe is
\begin{eqnarray}
t=\frac{1}{H_0}\int^{-\ln{(1+z)}}_{-\infty}\frac{dx}{h},\label{tage}
\end{eqnarray}
where $x=\ln{a}$ and $h=H/H_0$.
\begin{figure}[ht]
\begin{center}
\includegraphics[width=7cm]{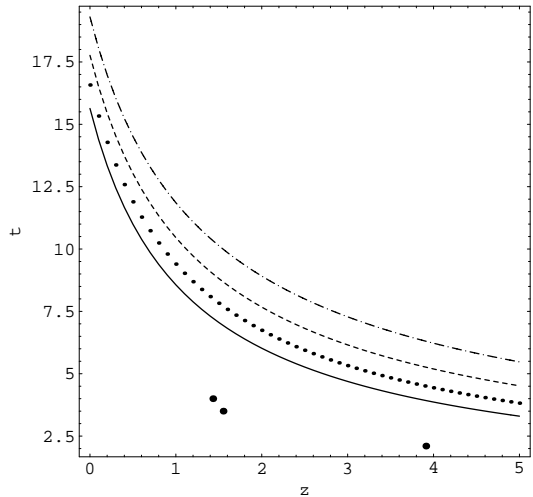}
\caption{The change of the age of Universe with the redshift $z$ for
different values of $\beta$. The solid line, dotted, dashed and
dotdashed denote the cases $\beta=0.61$, $0.62$, $0.63$ and $0.64$,
respectively. The large points denote the ages of the three old
objects LBDS $53W091$ $(z=1.55, t=3.5 Gyr)$, LBDS 53W069
($z=1.43,t=4.0 Gyr)$ and APM 08279+5255 $(z=3.91, t=2.1 Gyr)$. Here
we set $h_0=0.72$.}
\end{center}
\end{figure}
From the observations of the Hubble Space Telescope Key project, the
present Hubble parameter is constrained to be $H^{-1}_0
=9.776\;h^{-1}_0 Gyr$, where $0.64<h_0<0.80$. With Hubble parameter
$h_0= 0.72$ and fractional matter density $\Omega_{m0}=0.27$, the
$\Lambda$CDM model would give an age $t=1.6 Gyr$ at $z=3.91$, which
is smaller than the ages $2.1 Gyr$ inferred from old quasar APM
08279+5255. This is so-called ``high-$z$ cosmic age problem"
\cite{aget1}. Recently, it has attracted a lot of attention
\cite{aget1,aget2,aget3,aget4,aget5,aget6,aget7,aget8,aget9} and
various DE models are examined against these old galaxies and
quasars. So far, there is no DE model that can pass the examine, and
most of these models perform even poorer than $\Lambda$CDM model in
solving this problem. Some authors argued that the age problem can
be alleviated if one consider the interaction between dark energy
and dark matter \cite{aget9,aget10}. However, most of the interacted
forms are phenomenological because the natures of dark energy and
dark matter are still unclear at present. In this paper, we will
examine the second model by these old high redshift galaxies and
quasars. We set $h_0=0.72$ and draw the curves of age of the
universe. From Fig.(1) we find that that for the chosen values of
$\beta$ the age problem can be solved in this case. These results
imply that the effects of the torsion could help us to understand
more about our present Universe.

\section{Observational constraint}

We are now in position to use the observational data to fit the
seond models in the torsion cosmology in the Klauza-Klein theory.
Both the torsion parameters $\alpha$ and $\beta$ are determined by
minimizing $\chi^2=\chi^2_{SN}$. For the Type Ia SNe data, we use
the latest 307 Union SNIa data \cite{ob1} and define
\begin{eqnarray}
\chi^2_{SN}(\theta)=\sum^{307}_{i=1}\frac{[\mu_{obs}(z_i)-\mu(z_i)]^2}{\sigma^2_i},\label{chsn}
\end{eqnarray}
where $\mu_{obs}(z_i)$ and $\sigma_i$ are the observed value and the
total error for the supernova dataset, respectively. $\theta$ is the
parameter in the model. $\mu(z)$ is the theoretical distance modulus
which is given by
\begin{eqnarray}
\mu(z)=5\log_{10}{D(z)}+\mu_0,
\end{eqnarray}
where $\mu_0$ is a nuisance parameter. The luminosity of distance
$D(z)$ is
\begin{eqnarray}
D(z)=\frac{1+z}{H_0}\int^{z}_0\frac{dz'}{E(z'; \theta)},
\end{eqnarray}
and $E(z;\theta)$ is
\begin{eqnarray}
E(z;\theta)=\frac{H(z;\theta)}{H_0}.
\end{eqnarray}
As in \cite{ob2}, expanding the $\chi^2_{SN}$ of Eq. (\ref{chsn})
with respect to $\mu_0$, we have
\begin{eqnarray}
\chi^2_{SN}(\theta)=A(\theta)-2\mu_0B(\theta)+\mu^2_0C,\label{chsn1}
\end{eqnarray}
where
\begin{eqnarray}
A(\theta)&=&\sum^{307}_{i=1}\frac{[\mu_{obs}(z_i)-\mu(z_i,\mu_0=0)]^2}{\sigma^2_i},\nonumber\\
B(\theta)&=&\sum^{307}_{i=1}\frac{\mu_{obs}(z_i)-\mu(z_i,\mu_0=0)}{\sigma^2_i},\nonumber\\
C&=&\sum^{307}_{i=1}\frac{1}{\sigma^2_i}.
\end{eqnarray}
It is easy to see that when $\mu_0=B(\theta)/C$ the formula
(\ref{chsn1}) has a minimum
\begin{eqnarray}
\tilde{\chi}^2_{SN}(\theta)=A(\theta)-\frac{B(\theta)^2}{C}.
\end{eqnarray}
This means that
$\chi^2_{SN,min}(\theta)=\tilde{\chi}^2_{SN,min}(\theta)$. Since
$\tilde{\chi}^2_{SN,min}(\theta)$ is independent of $\mu_0$, we will
minimize $\tilde{\chi}^2_{SN,min}(\theta)$ rather than
$\chi^2_{SN}(\theta)$ in the our analysis.
\begin{figure}[ht]
\begin{center}
\includegraphics[width=7cm]{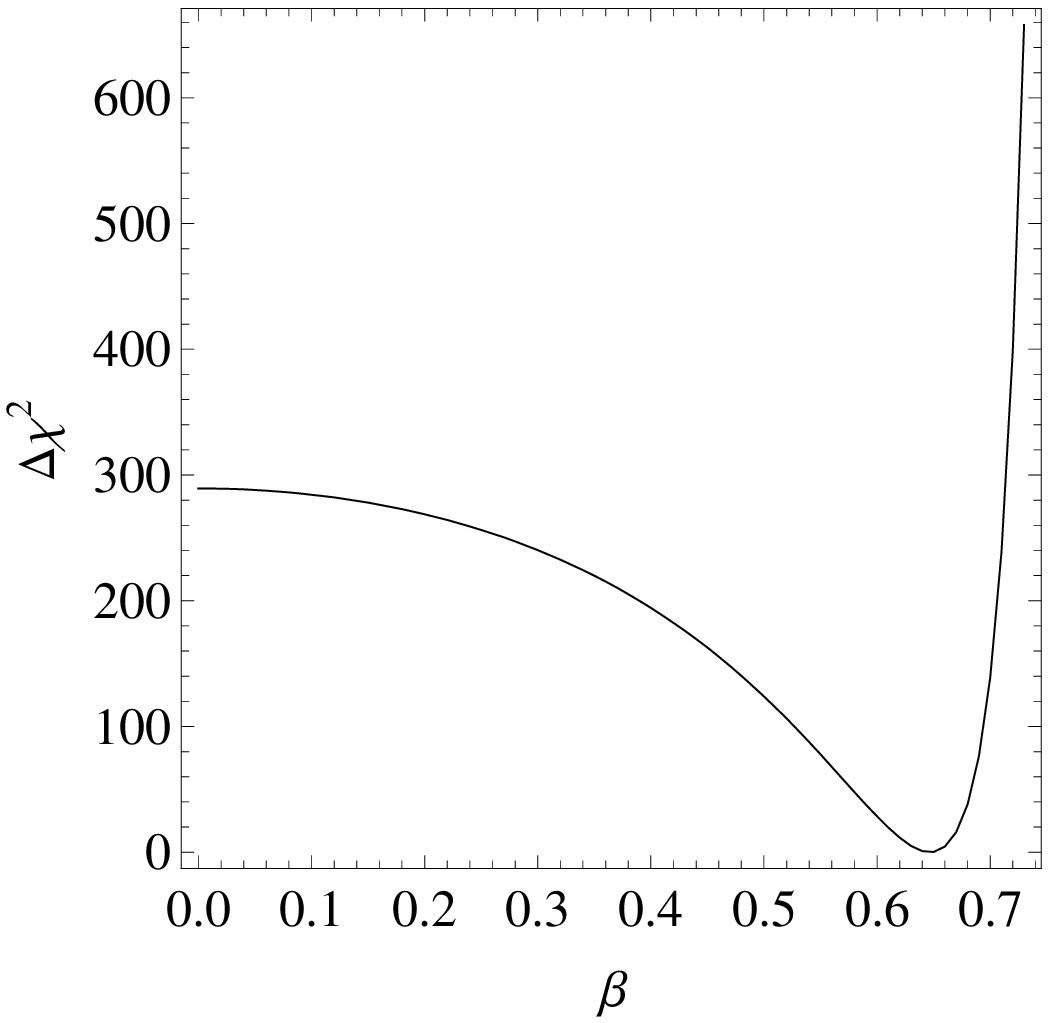}\;\;\;\;\;\includegraphics[width=7cm]{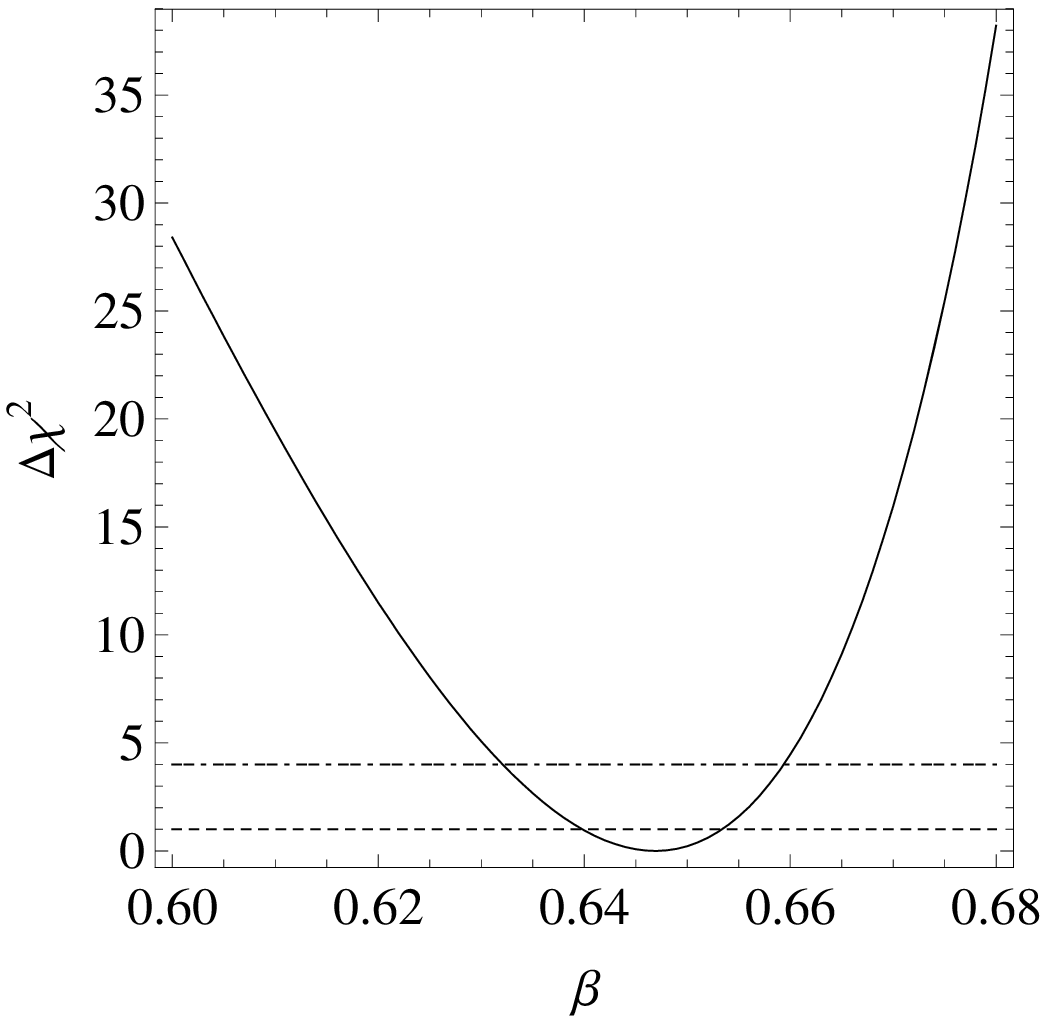}
\caption{The $\beta$-$\Delta\chi^2$ plane for the second model by
using the SNIa data. Here $\Delta\chi^2=\chi^2-\chi^2_{min}$ and
$\chi^2_{min}=318.84$. The dashed line and dash-dotted line in the
right figure denote $\Delta\chi^2=1$ and $\Delta\chi^2=4$,
corresponding to $68.3\%$ and $95.4\%$ C.L., respectively.}
\end{center}
\end{figure}

Now we present the numerical results of fitting the torsion
cosmology in Kaluza-Klein theory to the Type Ia SNe data. For the
second model $A_{\mu}=a(t)(0,\beta,\beta,\beta)$ and $\Phi^2=1$, we
obtain that $\chi^2_{min}=318.84$ and best-fit value of
$\beta=0.64689^{+0.00577}_{-0.00694}$ in the range of $1\sigma$,
which ensures  that the deceleration parameter $q<0$ and our present
Universe is accelerating. Moreover, we define the quantity $\Delta
\chi^2=\chi^2-\chi^2_{min}$ and then plot the $\beta$-$\Delta\chi^2$
plane in Fig. (2), which supports the best-fit value of $\beta$ we
obtained above. In the range of $2\sigma$, we find the model
parameter $\beta$ is $\beta\in (0.63232,\;0.65864)$. Comparing with
the chosen value of $\beta$ in Fig.(1), we find that the age problem
of three old objects can be solved in the torsion cosmology in the
Kaluza-Klein theory. This implies that the second model here we
considered is able to accommodate both the ages of the high redshift
objects and the SNIa measurements. However, it seems impossible for
the usual dark energy models. For example, although the $\Lambda$CDM
model is favored by the observational data, as our discussion
previously it is not free from the age problem of the oldest quasar
APM 08279+5255. The similar results are obtained in other dark
energy models. From these discussion, it is easy to conclude that
Kaluza-Klein type theories with torsion within proper constraints
will play an important role in cosmology.

\section{Summary}

In this paper we have studied the torsion cosmology model in the
Kaluza-Klein theory. We considered two simple models in which the
torsion vectors are $A_{\mu}=(\alpha,0,0,0)$ and
$A_{\mu}=a(t)^2(0,\beta,\beta,\beta)$, respectively. For the first
model, the accelerating expansion of the Universe can be not
explained without dark energy which is similar to that in the
standard cosmology. However, for the second model, we find that the
expansion of the universe can be accelerated without dark energy and
it is free of the age problem of the three old objects for
appropriated value of the model parameter $\beta$, which is
different from that in the Einstein cosmology. The constraints on
the second torsion cosmology models in Kaluza-Klein theory have been
studied by the analysis of SNIa data. The best-fit value of the
parameter in the model is $\beta=0.6468$. It is worth noting that in
the $2\sigma$ uncertainty the second model is able to accommodate
both the ages of the high redshift objects and the SNIa
measurements. These means that modified gravity theories obtained by
Kaluza-Klein theories with torsion will play an important role in
cosmological models.

\acknowledgments We thank the referee for his/her quite useful and
helpful comments and suggestions, which help deepen our
understanding of the torsion cosmology in Kaluza-Klein theory. This
work was partially supported by the National Natural Science
Foundation of China under Grant No.10875041 and the construct
program of key disciplines in Hunan Province. J. L. Jing's work was
partially supported by the National Natural Science Foundation of
China under Grant No.10675045 and No.10875040; and the Hunan
Provincial Natural Science Foundation of China under Grant
No.08JJ3010.

\end{document}